\documentclass[prd,twocolumn,showpacs,amsmath,amssymb]{revtex4}
\usepackage{graphicx}
\begin{document}
\title{Friedmann cosmology with a generalized equation of state and bulk viscosity}
\author{Xin-He Meng}
\email{xhm@nankai.edu.cn}
\author{Jie Ren}
\email{jrenphysics@hotmail.com}
\author{Ming-Guang Hu}
\email{hu_mingguang@hotmail.com} \affiliation{Department of
physics, Nankai University, Tianjin 300071, China}
\begin{abstract}
The universe media is considered as a non-perfect fluid with bulk
viscosity and described by a more general equation of state. We
assume the bulk viscosity is a linear combination of the two
terms: one is constant, and the other is proportional to the
scalar expansion $\theta=3\dot{a}/a$. The equation of state is
described as $p=(\gamma-1)\rho+p_0$, where $p_0$ is a parameter.
This model can be used to explain the dark energy dominated
universe. Different choices of the parameters may lead to three
kinds of fates of the cosmological evolution: no future
singularity, big rip, or Type III singularity of Ref.~[S. Nojiri,
S.D. Odintsov, and S. Tsujikawa, Phys. Rev. D \textbf{71}, 063004
(2005)].
\end{abstract}
\maketitle

\section{Introduction}
The cosmological observations indicate that the expansion of our
universe accelerates \cite{bah99}. Recently£¬ lots of work on
extended gravity \cite{mw1}£¬such as modifying equation of state
or by introducing the so called dark energy is to explain the
cosmic acceleration expansion observed. To overcome the drawback
of hydrodynamical instability, a linear equation of state of a
more general form, $p=\alpha(\rho-p_0)$ is proposed \cite{bab05},
and this form incorporated into cosmological model can describe
the hydrodynamically stable dark energy behaviors.

The observations also indicate that the universe media is not a
perfect fluid \cite{jaf05} and the viscosity is concerned in the
evolution of the universe \cite{bre05a,bre05b,cat05}. In the
standard cosmological model, if the equation of state parameter
$\omega$ is less than -1, the universe shows the future finite
singularity called Big Rip \cite{cal03,noj05}. Several ideas are
proposed to prevent the big rip singularity, like by introducing
quantum effects terms in the action.

In this paper, we show that the Friedmann equations can be solved
with both a more general equation of state and bulk viscosity
detailed as follows. The equation of state is
\begin{equation}
p=(\gamma-1)\rho+p_0,
\end{equation}
where $p_0$  and $\gamma$ are two parameters. The bulk viscosity
is expressed as
\begin{equation}
\zeta=\zeta_0+\zeta_1\frac{\dot{a}}{a}.
\end{equation}
where $\zeta_0$ and $\zeta_1$ are two constants conventionally.
The $\omega=p/\rho$ is constrained as $-1.38<\omega<-0.82$
\cite{mel03} by present observation data, so the inequality in our
case should be
\begin{equation}
-1.38<\gamma-1+\frac{p_0}{\rho}<-0.82.
\end{equation}
The parameter $p_0$ can be positive (attractive force) or negative
(repulsive force), and conventionally $\zeta_0$ and $\zeta_1$ are
regarded as positive. To choose the parameters properly, it can
prevent the Big Rip problem or some kind of singularity for the
cosmology model, like in the phantom energy phase, as shown below.

This paper is organized as follows. In Sec. II we describe our
model and give out the exact solution. In Sec. III we consider
some special cases of the solution. In Sec. IV we discuss the
acceleration phase and the future singularities in this model, and
in the last section (Sec. V) we summarize our conclusions.

\section{Model and calculations}
We consider the Friedamnn-Roberson-Walker metric in the flat space
geometry (k=0)
\begin{equation}
ds^2=-dt^2+a(t)^2(dr^2+r^2d\Omega^2),
\end{equation}
and assume that the cosmic fluid possesses a bulk viscosity
$\zeta$. The energy-momentum tensor can be written as
\begin{equation}
T_{\mu\nu}=\rho U_\mu U_\nu+(p-\zeta\theta)H_{\mu\nu},
\end{equation}
where in comoving coordinates $U^\mu=(1,0)$,
$\theta=U^\mu_{;\mu}=3\dot{a}/a$, and $H_{\mu\nu}=g_{\mu\nu}+U_\mu
U_\nu$ \cite{bre02}. By defining the effective pressure as
$\tilde{p}=p-\zeta\theta$ and from the Einstein equation
$R_{\mu\nu}-\frac{1}{2}g_{\mu\nu}R=8\pi GT_{\mu\nu}$, we obtain
the Friedmann equations
\begin{subequations}
\begin{eqnarray}
\frac{\dot{a}^2}{a^2} &=& \frac{8\pi G}{3}\rho\label{eq1},\\
\frac{\ddot{a}}{a} &=& -\frac{4\pi
G}{3}(\rho+3\tilde{p})\label{eq2}.
\end{eqnarray}
\end{subequations}
The conservation equation for energy, $T^{0\nu}_{;\nu}$, yields
\begin{equation}
\dot{\rho}+(\rho+\tilde{p})\theta=0.
\end{equation}

Using the equation of state to eliminate $\rho$ and $p$, we obtain
the equation which determines the scale factor $a(t)$
\begin{equation}
\frac{\ddot{a}}{a}=-\frac{3\tilde{\gamma}-2}{2}\frac{\dot{a}^2}{a^2}+12\pi
G\zeta_0\frac{\dot{a}}{a}-4\pi Gp_0\label{eq3},
\end{equation}
where the effective equation of state parameter is shifted from
the original one as
\begin{equation}
\tilde{\gamma}=\gamma-8\pi G\zeta_1.
\end{equation}
So we can see that the equivalent effect of the second term in
$\zeta$ is to change the parameter $\gamma$ to $\tilde{\gamma}$ in
the equation of state. As shown in Ref.~\cite{bre05a}, the barrier
$\omega=-1$ between the quintessence region ($\omega>-1$) and the
phantom region ($\omega<-1$) can be crossed, as a consequence of
the bulk viscosity available.

Since the dimension of the two terms $12\pi G\zeta_0$ and $-4\pi
Gp_0$ is $[\rm{time}]^{-1}$ and $[\rm{time}]^{-2}$, respectively,
we define
\begin{eqnarray}
12\pi G\zeta_0 &=& \frac{1}{T_1},\\
-4\pi Gp_0 &=& \frac{1}{T_2^2},
\end{eqnarray}
then Eq.~(\ref{eq3}) becomes
\begin{equation}
\frac{\ddot{a}}{a}=-\frac{3\tilde{\gamma}-2}{2}\frac{\dot{a}^2}{a^2}
+\frac{1}{T_1}\frac{\dot{a}}{a}+\frac{1}{T_2^2}.\label{eq:main}
\end{equation}
Here $T_1$ and $T_2$ are criteria to determine whether we should
concern the $\zeta$ and $p_0$. If $T_1>>t, the cosmic time scale$,
the effect of $\zeta$ can be neglected, and if $T_2>>t$, the
effect of $p_0$ can be neglected likewise.

Concerning the initial conditions of $a(t_0)=a_0$ and
$\theta(t_0)=\theta_0$, if $\tilde{\gamma}\neq 0$, the solution
can be obtained as
\begin{widetext}
\begin{equation}
a(t)=a_0\left\{\frac{1}{2}\left(1+\tilde{\gamma}\theta_0
T-\frac{T}{T_1}\right){\rm{exp}}\left[\frac{t-t_0}{2}\left(\frac{1}{T}
+\frac{1}{T_1}\right)\right]+\frac{1}{2}\left(1-\tilde{\gamma}\theta_0
T+\frac{T}{T_1}\right){\rm{exp}}\left[-\frac{t-t_0}{2}\left(\frac{1}{T}
-\frac{1}{T_1}\right)\right]\right\}^{2/3\tilde{\gamma}}.\label{eq4}
\end{equation}
And we obtain directly
\begin{equation}
\frac{\dot{a}}{a}=\frac{1}{3\tilde{\gamma}}\frac{(1+\tilde{\gamma}\theta_0
T-\frac{T}{T_1})(\frac{1}{T}+\frac{1}{T_1}){\rm{exp}}(\frac{t-t_0}{T})
-(1-\tilde{\gamma}\theta_0
T+\frac{T}{T_1})(\frac{1}{T}-\frac{1}{T_1})}{(1+\tilde{\gamma}\theta_0
T-\frac{T}{T_1}){\rm{exp}}(\frac{t-t_0}{T})+(1-\tilde{\gamma}\theta_0
T+\frac{T}{T_1})}.\label{eq5}
\end{equation}
\end{widetext}
Here we define
\begin{equation}
T=\frac{T_1}{\sqrt{1+6\tilde{\gamma}(T_1/T_2)^2}}.
\end{equation}
We can see that when $T_2\rightarrow\infty$, $T=T_1$; when
$T_1\rightarrow\infty$, $T=T_2/\sqrt{6\tilde{\gamma}}$.

\section{$\tilde{\gamma}=0$ and special cases}
For $\tilde{\gamma}=0$, we should use the mathematical
L'Hospital's rule to calculate the limit of Eq.~(\ref{eq4})
rigously and note
\begin{equation}
\lim_{\tilde{\gamma}\to
0}\frac{dT}{d\tilde{\gamma}}=-\frac{3T_1^3}{T_2^2}.
\end{equation}
The limit of solution $a(t)$ when $\tilde{\gamma}\to 0$ is
\begin{eqnarray}
a(t)=a_0{\rm{exp}}[\left(\frac{1}{3}\theta_0T_1+\frac{T_1^2}{T_2^2}
\right)\left(e^{(t-t_0)/T_1}-1\right)\nonumber\\
-\frac{T_1(t-t_0)}{T_2^2}].\label{eq:gamma}
\end{eqnarray}
Directly solving Eq.~(\ref{eq:main})
\begin{equation}
\frac{\ddot{a}}{a}=\frac{\dot{a}^2}{a^2}+\frac{1}{T_1}\frac{\dot{a}}{a}
+\frac{1}{T_2}
\end{equation}
gives the same result. So Eq.~(\ref{eq4}) is consistent for
$\tilde{\gamma}$ crossing zero. The $T_2\to\infty$ limit of
Eq.~(\ref{eq:gamma}) is
\begin{equation}
a(t)=a_0{\rm{exp}}\left[\frac{1}{3}\theta_0T_1\left(e^{(t-t_0)/T_1}-1\right)\right].
\end{equation}
and the $T_1\to\infty$ limit of Eq.~(\ref{eq:gamma}) is
\begin{equation}
a(t)=a_0{\rm{exp}}\left[\frac{1}{3}\theta_0(t-t_0)+\frac{(t-t_0)^2}{2T_2^2}\right].
\end{equation}
These two special limits are also consistent with directly solving
Eq.~(\ref{eq:main}), by checking.

Let us discuss two special cases in the following. When the
constant term in the equation of state is not concerned, i.e.
$T_2\rightarrow\infty$,
\begin{equation}
a(t)=a_0\left[1+\frac{1}{2}\tilde{\gamma}\theta_0
T\left(e^{(t-t_0)/T}+1\right)\right]^{2/3\tilde{\gamma}}.
\end{equation}
and when the constant term in the bulk viscosity is not concerned,
i.e. $T_1\rightarrow\infty$,
\begin{equation}
a(t)=a_0\left({\rm{cosh}}\frac{t-t_0}{2T}+\tilde{\gamma}\theta_0
T{\rm{sinh}}\frac{t-t_0}{2T}\right)^{2/3\tilde{\gamma}}.
\end{equation}
When $T\to\infty$, the two cases become
\begin{equation}
a(t)=a_0\left[1+\frac{1}{2}\tilde{\gamma}\theta_0(t-t_0)\right]^{2/3\tilde{\gamma}}.
\end{equation}
For $\tilde{\gamma}\to 0$, the limit case is
\begin{equation}
a(t)=a_0 e^{\theta_0(t-t_0)/3},
\end{equation}
which corresponds to the de Sitter universe with accelerating
cosmic expansion.

Additional notions: Eqs.~(\ref{eq1}) and (\ref{eq2}) can be
rewritten as
\begin{subequations}
\begin{eqnarray}
H^2 &=& \frac{8\pi G}{3}\rho,\\
\dot{H} &=& -4\pi G(\tilde{p}+\rho).
\end{eqnarray}
\end{subequations}
From these equations, the relation among viscosity, the scalar
factor $a$ and Hubble parameter $H$ is
\begin{equation}
aH\frac{dH}{da}=-\frac{3\tilde{\gamma}}{2}H^2+12\pi G\zeta H-6\pi
Gp_0.
\end{equation}
which reflects the viscosity functions for dark energy and matter
dominated universe evolution.

\section{Acceleration and big rip}
If the universe accelerates, then mathematically
\begin{equation}
\frac{\ddot{a}}{a}>0.
\end{equation}
From Eq.~(\ref{eq3}), we can qualitatively see that the bulk
viscosity and a negative density $p_0$ can cause the universe to
accelerate. Since the expression of $\ddot{a}/a$ is too
complicated in this situation, now we only discuss a special case,
with $p_0=0$. Here $\ddot{a}/a>0$ yields
\begin{equation}
\omega=\gamma-1>\frac{2}{3}e^{(t-t_0)/T}-1+\frac{2}{\theta_0
T}+8\pi G\zeta_1.\label{eq6}
\end{equation}
As we know, if the bulk viscosity is zero as in the standard
Friedammn£­Robertson-Walker cosmology model, an accelerating
expansion universe corresponds to $\omega<-1/3$. Inequality
(\ref{eq6}) tells us that if the bulk viscosity is large enough,
the universe expansion can accelerate even if $\omega>-1/3$.

According to \cite{noj05}, the future singularities can be
classified in the following way:
\begin{itemize}
\item  Type I (``Big Rip''): For $t \to t_s$, $a \to \infty$,
$\rho \to \infty$ and $|p| \to \infty$
\item  Type II (``sudden''): For $t \to t_s$, $a \to a_s$,
$\rho \to \rho_s$ and $|p| \to \infty$
\item  Type III: For $t \to t_s$, $a \to a_s$,
$\rho \to \infty$ and $|p| \to \infty$
\item  Type IV: For $t \to t_s$, $a \to a_s$,
$\rho \to 0$, $|p| \to 0$ and higher derivatives of $H$ diverge.
\end{itemize}
In this paper, $\rho\to\infty$ means $p\to\infty$ (we assume
$\gamma\neq 1$ generally). In the following we show that different
choices of the parameters may lead to three fates of the universe
evolution: no future singularity, big rip, or the Type III
singularity.

\subsection{$\tilde{\gamma}<0$}
From Eq.~(\ref{eq1}), we see
\begin{equation}
\sqrt{\rho}\propto\frac{\dot{a}}{a}.
\end{equation}
If the denominator of Eq.~(\ref{eq5}) is zero,
\begin{equation}
\left(1+\tilde{\gamma}\theta_0
T-\frac{T}{T_1}\right){\rm{exp}}\left(\frac{t-t_0}{T}\right)
+1-\tilde{\gamma}\theta_0 T+\frac{T}{T_1}=0.
\end{equation}
then $a\to\infty$, $\rho\to\infty$, so the big rip occurs. The
solution for $t$ is
\begin{equation}
t_{\rm{s}}=T{\rm{ln}}\left(-\frac{1-\tilde{\gamma}\theta_0
T+\frac{T}{T_1}}{1+\tilde{\gamma}\theta_0
T-\frac{T}{T_1}}\right)+t_0
\end{equation}
If we want to prevent the big rip, there should be no real solution
for $t>t_0$, so
\begin{equation}
-\frac{1-\tilde{\gamma}\theta_0
T+\frac{T}{T_1}}{1+\tilde{\gamma}\theta_0 T-\frac{T}{T_1}}<1.
\end{equation}
The inequality is equivalent to
\begin{equation}
1+\tilde{\gamma}\theta_0 T-\frac{T}{T_1}>0.\label{eq7}
\end{equation}
The above inequality can be satisfied in some conditions for the
phantom energy, so the big rip will not occur. Furthermore, we can
see that even if the dark energy is in the quintessence region,
there also can be future singularity. For example, if
$\tilde{\gamma}>0$ and $p_0<0$, it is possible that inequality
~(\ref{eq7}) is not satisfied, so the future singularity may
occur, which will be discussed in the next subsection below.

The more explicit form of inequality ~(\ref{eq7}) is
\begin{equation}
1+\frac{\tilde{\gamma}\theta_0
T_1}{\sqrt{1+6\tilde{\gamma}(T_1/T_2)^2}}-
\frac{1}{\sqrt{1+6\tilde{\gamma}(T_1/T_2)^2}}>0.
\end{equation}
From this inequality, we obtain that

(i) $p_0<0$, i.e. $T_2^2>0$: inequality ~(\ref{eq7}) is always
unsatisfied, so there will be a big rip at time $t_{\rm{s}}$.

(ii) $p_0>0$, i.e. $T_2^2<0$: inequality ~(\ref{eq7}) is not
always unsatisfied. If it is unsatisfied, there will be a big rip
at time $t_{\rm{s}}$; if it is satisfied, there is no future
singularity.

\subsection{$\tilde{\gamma}>0$}
If the denominator of  inequality ~(\ref{eq5}) is zero, then $a\to
0$, $\rho\to \infty$, so the Type III singularity occurs.
Following the same steps as before, we obtain that

(i) $p_0<0$, i.e. $T_2^2>0$: inequality ~(\ref{eq7}) is always
satisfied, so there is no future singularity.

(ii) $p_0>0$, i.e. $T_2^2<0$: inequality ~(\ref{eq7}) is not
always satisfied. If it is satisfied, there is no future
singularity; if it is unsatisfied, there will be Type III
singularity at time $t_{\rm{s}}$.

\subsection{$\tilde{\gamma}=0$}
For the case $\tilde{\gamma}=0$,
\begin{equation}
\frac{\dot{a}}{a}=\frac{1}{3}\theta_0e^{(t-t_0)/T_1}
+\frac{T_1}{T_2^2}\left(e^{(t-t_0)/T_1}-1\right).
\end{equation}
So there is no future singularity in this case.

To illustrate the parameters in the general solution Eq
~(\ref{eq4}) more clearly, we draw some graphics in Fig.~1-4. The
initial condition is \cite{bre02} $t_0=1000{\rm{s}}$,
$\theta_0=1.5\times 10^{-3}{\rm{s^{-1}}}$. At this time, the bulk
viscosity $\zeta=7.0\times 10^{-3}{\rm{g/cm\cdot s}}$, the
corresponding $T_1=5.1\times 10^{28}{\rm{s}}$. We assume
$\zeta_1=0$ for simplicity.
\begin{figure}[]
\includegraphics{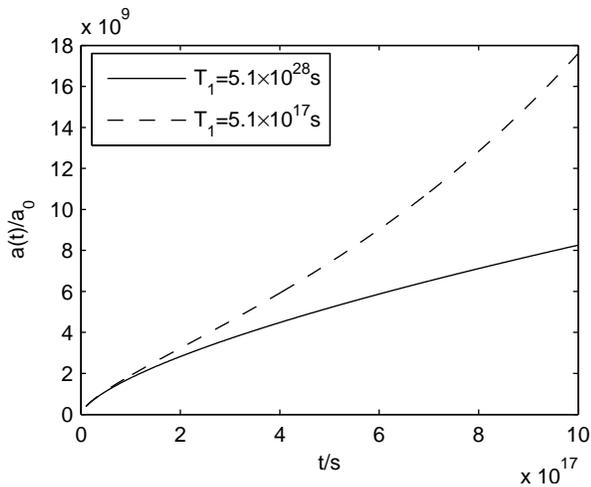}
\caption{\label{fig1} $p_0=0$, $\gamma=1$, the dash line
corresponds to $T_1=5.1\times 10^{28}$s, and another curve
corresponds to $T_1=5.1\times 10^{17}$s.}
\end{figure}
\begin{figure}[]
\includegraphics{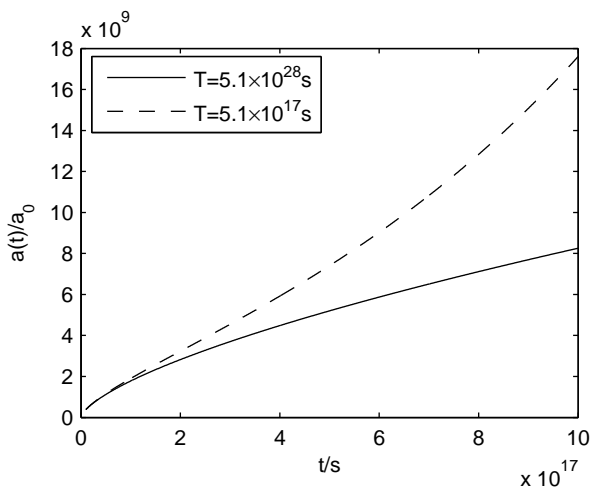}
\caption{\label{fig2} $\zeta=0$, $p_0=0$, the dash line
corresponds to $\gamma=0.181$, another curve  corresponds to
$\gamma=0.18$ case.}
\end{figure}\begin{figure}[]
\includegraphics{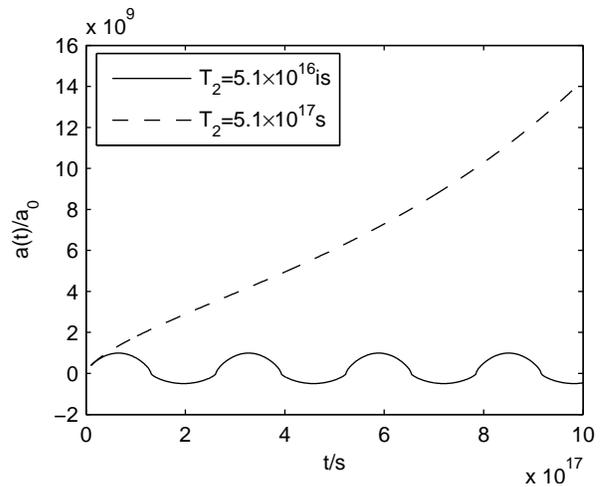}
\caption{\label{fig3} $\zeta=0$, $\gamma=1$, the dash line
corresponds to $T_2=5.1\times 10^{16}i$s, while another curve
corresponds to $T_2=5.1\times 10^{17}$s case.}
\end{figure}\begin{figure}[]
\includegraphics{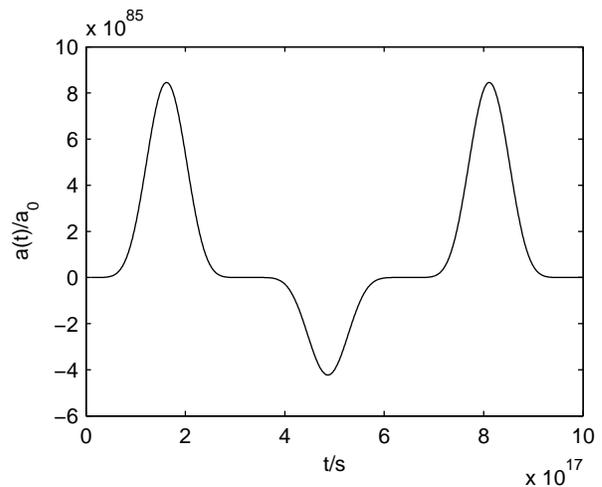}
\caption{\label{fig4} $\zeta=0$, $T_2=5.1\times 10^{16}i$s,
$\gamma=0.1$.}
\end{figure}

\section{Conclusion}
In conclusion, we have solved the Friedmann equations with both a
more general equation of state and bulk viscosity, and discussed
the acceleration expansion of the universe evolution and the
future singularities for this model. Compared with the standard
model of cosmology, this model has had three additional
parameters, $\zeta_0$, $\zeta_1$ and $p_0$: choices of $\zeta_0$
and negative $p_0$ can cause the universe accelerate; $\zeta_1$
can drive the cosmic fluid from the quintessence region to the
phantom one \cite{bre05a}, and positive $p_0$ may both prevent the
big rip for phantom phase and lead to the Type III singularity of
Ref.~\cite{noj05} for the quintessence phase. The relation between
the choices of parameters and the future singularities of the
cosmological evolution in this extended model is summarized as in
the following table and we expect more detail investigations on
viscosity effects to be carried out.

\begin{tabular}{c|c|c|c}
\hline
\multicolumn{3}{c|}{Parameters} & Future singularity\\
\multicolumn{3}{c|}{} & (at $t\to t_s$)\\
\hline
$\tilde{\gamma}<0$ & \multicolumn{2}{|c|}{$p_0<0$} & $a\to\infty, \rho\to\infty$\\
\cline{2-4}
& $p_0>0$ & $1+\tilde{\gamma}\theta_0 T-T/T_1>0$ & No\\
\cline{3-4} & & $1+\tilde{\gamma}\theta_0 T-T/T_1<0$ & $a\to\infty,
\rho\to\infty$\\
\hline
\multicolumn{3}{c|}{$\tilde{\gamma}=0$} & No\\
\hline
$\tilde{\gamma}>0$ & \multicolumn{2}{|c|}{$p_0<0$} & No\\
\cline{2-4}
& $p_0>0$ & $1+\tilde{\gamma}\theta_0 T-T/T_1>0$ & No\\
\cline{3-4} & & $1+\tilde{\gamma}\theta_0 T-T/T_1<0$ & $a\to 0, \rho\to\infty$\\
\hline
\end{tabular}

{Aknowledgement}

We thank Prof. I. Brevik for reading the manuscript with helpful
comments and Profs. S.D.Odintsov and Lewis H.Ryder for lots of
interesting discussions. This work is partly supported by NSF and
Doctoral Foundation of China.

\end{document}